\documentclass[12pt,preprint]{aastex}

\begin{document}
\title{
The Amazing Old Nova Q Cygni: A Far Ultraviolet Synthetic Spectral Analysis}
\author{
Craig Kolobow and Edward Sion}
\affil{Department of Astronomy and Astrophysics
Villanova University,
Villanova, PA 19087\\craig.kolobow@villanova.edu, edward.sion@villanova.edu}

\begin{abstract}
Q Cygni (Nova Cygni 1876) is the third oldest old novae (after 
WY Sge and V841 Oph) with a long orbital period of 10.08 hours and 
spectroscopic peculiarities in the optical including the presence 
of variable wind outflow revealed by optical P Cygni profiles in 
the He\,{\sc i} lines and H$ \alpha \beta $ (Kafka et al. 2003). We have carried 
out a synthetic spectral analysis of a far ultraviolet IUE archival spectrum of 
Q Cygni using our optically thick, steady state, accretion disk models and model white dwarf 
photospheres. We find that the accretion light of a luminous accretion disk dominates the
FUV flux of the hot component with a rate of accretion $2 -3\times 10^{-9}$ M$_{\sun}$/yr.
We find that Q Cygni lies at a distance of $741 \pm 110 pc$ . The implications of our results 
for theoretical predictions for old novae are presented. 
\end{abstract}

\section{ Introduction}

Q Cygni (Nova Cygni 1876) is one of the oldest old novae with a long orbital period of 10.08 hours and 
spectroscopic peculiarities in the optical including the presence of variable wind outflow revealed by 
optical P Cygni profiles in the He I lines and 
H$\alpha  $ (Kafka et al. 2003). Their time-resolved optical spectra revealed pronounced P Cygni 
profiles in the He I triplet lines at 5876\AA\ and 7065\AA\ and in H$\alpha$ indicating the presence 
of an outflow with velocity reaching 1500 km sec$^{-1}$. The wind outflow is highly variable with time and 
with orbital phase. It is rare to find outflow signatures of CV winds in the optical. 

Selvelli (2004) summarized the UV spectral properties of 18 old novae 
observed with IUE. He found that their de-reddened continuum energy 
distribution is well described by either a power-law distribution 
$F_{\lambda} \sim \lambda^{- \alpha}$ with $\alpha$ in the range 0.3 to 2.5 or by a single 
black-body distribution with T in the range 15 000 - 38 000 K. 
If one assume that the UV luminosities come entirely from an accretion disk, then 
one obtains that the "average" disk luminosity disk is about 
$20 L_{\odot}$ and that the "average" accretion rate $\dot{M}$ 
is $\sim 1 \times 10^{-8}M_{\odot}~$yr$^{-1}$. Much remains to be 
learned about old novae including how well their optical and UV behavior as they enter quiescence agrees with predictions of hibernation theory. Q Cyg has been previously thought to be entering a state of hibernation, where the secondary star detaches from the Roche lobe.  This has been supported from ground based estimates that the accretion rate has been gradually declining since the 1876 outburst (Schaefer 2010). For Q Cygni, there has not been a far UV synthetic spectroscopic analysis of this system  using optically thick, steady state, accretion disk models and model white dwarf photospheres. We report the results of such a spectroscopic analysis and compare the physical paranmeters we derive with those of other old novae.

\section{Far Ultraviolet Spectroscopic Observations}

The IUE spectroscopic observations of Q Cygni were carried out on 01-01-1989 starting at 10:51:36 UT. The spectrum SWP35239 had an exposure time of 14,160 seconds at low dispersion through the large aperture. An LWP spectrum (LWP14754) was also obtained which is quite noisy but indicates the expected absorption trough at 2200\AA\ due to interstellar absorption. The observed reddened spectrum has a well-exposed continuum slightly rising with average flux level of $\sim 7 \times 10^{-15}$ ergs/cm${^2}$/s/\AA\ over the SWP wavelength range of 1170\AA\ to 2000\AA\. There appear to be hints of emission features but nothing that can be clearly identified due to the low quality and signal to noise of the spectrum. The Lyman $\alpha$  profile is filled in with emission due to geocoronal contamination of the large aperture spectrum. What appears to be a broad absoption trough centered near 1500\AA\ could be pseudo-absorption due to two neighboring emission regions. There do appear to be absorption wings at Lyman $\alpha$. We de-reddened the spectrum with$ E(B-V) = 0.20$, a reasonable value for its line of sight. 

\section{ Method of Model Fitting}

We adopted model accretion disks from the optically thick disk model grid of Wade and Hubeny (1998). In these accretion disk models, the innermost disk radius, $R_{in}$, is fixed at a fractional white dwarf radius of $x = R_{in}/R_{wd} = 1.05$. The outermost disk radius, $R_{out}$, was chosen so that T$_{eff}(R_{out})$  is near 10,000K since disk annuli beyond this point, which are cooler zones with larger radii, would provide only a very small contribution to the mid and far UV disk flux, particularly the SWP FUV band pass. The mass transfer rate is assumed to be the same for all radii. For a given spectrum, we carry out fits for every combination of \.{M}, inclination and white dwarf mass in the Wade \&  Hubeny (1998) library. The values of $ i$ are 18, 41, 60, 75 and 81. The range of accretion rates covers $-10.5 < \log M < -8.0 $ in steps of 0.5 in the log and five different values of the white dwarf mass, namely, 0.4, 0.55, 0.8, 1.0, and 1.2 M$_{\sun}$. The emission line regions that were masked out were 1195 \AA - 1245 \AA, 1280\AA - 1290\AA, and 1650 \AA - 1670 \AA.  For the disk models, we selected all models with inclination angle $ i = 41, 18$, M$_{wd} = 0.35, 0.55, 0.80, 1.03,  1.21$  M$_{\sun}$ and -$\log (\dot{M}) = 8.0, 8.5, 9.0, 9.5, 10.0, 10.5$. The WD models used temperatures of 10,000K to 80,000K in steps of 1,000K and log $g =7.0,7.5,8.0,8.5,9.0$.

\section{ Model Fitting Results}

In preparation for the model fitting we masked out possible  
emission line regions at 1195\AA - 1245\AA,  1280\AA - 1290\AA\ and 1650\AA - 1670\AA.
Our fitting procedure to the de-reddened IUE spectrum, using our $\chi^{2}$ minimization routine called IUEFIT, consisted of 
first using models of hot white dwarfs only, then fitting the spectrum with accretion disk models only and finally, fitting combinations of white dwarf and accretion models, if a statistically significant improvement resulted.
We combined white dwarf models and accretion disk models using a $\chi^{2}$ minimization routine called DISKFIT. Using these routines. the best-fitting WD-only, disk-only and composite white dwarf plus disk model is determined based upon the minimum $\chi^{2}$ value achieved, visual inspection ($\chi$-by eye) of the model, consistency with the continuum slope and Ly$\alpha$  region, and the consistency of the scale factor-derived distance with published distance estimates or trigonometric parallax distance.

The results of our fitting is given in Table 1 where we list by column: (1) the type of model used, (2) the T$_{eff}$ of the WD model; (3) the accretion rate; (4)
the inclination; (5) the WD mass of the model; (6) the $\chi^{2}$ value; (7) the 
scale factor of the fit; and (8) the distance implied by the model fit.

\begin{deluxetable}{lccccccc}
\tablecolumns{8}
\tablewidth{0pc}
\tablecaption{Model Fitting Results}
\tablehead{
\colhead{Model} 
& \colhead{T$_{eff}$K} 
&\colhead{$\dot{M}(M_{\sun})$/yr}
&\colhead{$i$(deg)}
&\colhead{WDmass($M_{\sun}$)}
&\colhead{$\chi^{2}$}
&\colhead{Scale Factor}
&\colhead{Distance(pc)}
}
\startdata
WD   &   31000  &     ----      &     ---    &   0.8     &     2.51 & $ 4.2\times 10^{-3}$  &  169\\
Disk   &  ---      & $3\times 10^{-9}$    &   41      &  0.55   &      2.30 &  $1.82\times10^{-2}$  &  741 \\

WD+Disk& 31000 &     $2.82\times 10^{-9}$  &  41      &  0.55        & 2.29  & $1.82\times 10^{-2}$  & 741\\
\enddata
\end{deluxetable}

The best fitting WD-only model had T$_{eff} = 31000$K, Log g$= 8.5$ and is displayed in Fig.1 but the distance implied by the fit (169 pc) 
is unrealistically too close so this solution was rejected. Nevertheless, the FUV energy distribution of a 31000K WD is within the range of the 
black body temperatures characterizing the observed energy distribution.

The best fitting accretion disk, displayed in Fig.2, had
a slightly smaller $\chi^{2}$ and yielded a reasonable distance of $741 pc \pm 110$. We also tried a combination WD + disk model
with a best fit obtained with the WD contributing only 10\% of the FUV flux and the disk contributing 90\%. This combination fit is shown in Fig.3.
Since there was no statistically significant improvement over the fit with an accretion disk alone, we rejected this combination
and adopted the accretion disk-only fit as the most reasonable solution for Q Cygni.

\section{Summary}
The accretion rate that we derive for Q Cygni, $3\times 10^{-9}$ M$_{\sun}$/yr)
is below the average accretion rate derived in the FUV for old novae but still considerable
higher than the theoretically expected low accretion rates needed to power a strong nova explosion.

Our distance of 741 pc is much closer than the distance used by Kafka et al. (2003) which was based upon on the Na D 
absorption lines in their optical spectra and yielded a distance of 
the order of 3 kpc, about a factor of 4 further than our model-derived distance. On the other hand, the distance to Q Cygni using the 
$t_3$ method of Duerbeck (1981) with $t_3 = 11$ days (a very fast nova) yields a distance d = 1.9 kpc. 
For an old nova at a distance of 750 pc, with V magnitude $\sim$ 15 (from the photometry
of Kafka et al. 2003), and a visual absorption $A_v = 0.8$ mag, the resulting absolute magnitude is 4.85. This is in very good agreement
agreement with the absolute magnitude of classical novae in quiescence
which was about 4.2 (see Patterson 1984). A 3 kpc distance
measurement yields an absolute magnitude of 1.8,  which would make Q Cygni a
super-luminous old nova, and a correspondingly much higher accretion rate.

In summary, the hot component is overwhelmingly dominated by accretion light from a luminous disk surrounding the white dwarf. 
The distance that we derived from the modeling (740 pc) is not unreasonable and the accretion rate we derive ($2 -3\times 10^{-9}$ M$_{\sun}$/yr) in the FUV 
indicates that the mass transfer rate is still fairly high 112 years after the nova explosion in 1876.

It is a pleasure to thank Stella Kafka and the referee, Michael Friedjung, for helpful comments.
We gratefully acknowledge the support of this work by NSF grant 0807892 to Villanova University and partial summer 
undergraduate support from the NASA Delaware Space Grant Consortium.

\section{References}

Bruch, A.,  Engel, 1994, AA Supp., 104, 79

Duerbeck, H.1981, PASP, 93, 165

Kafka, S., et al. 2003, AJ, 126, 1422

Patterson 1984, ApJS, 54, 443

Schaefer, B.2010, private communication

Selvelli, P.2004, Baltic Astr.13, 93

Wade, R.A. \& Hubeny, I. 1998, ApJ, 509, 350.


\clearpage
\begin{figure}
\epsscale{.90} 
\plotone{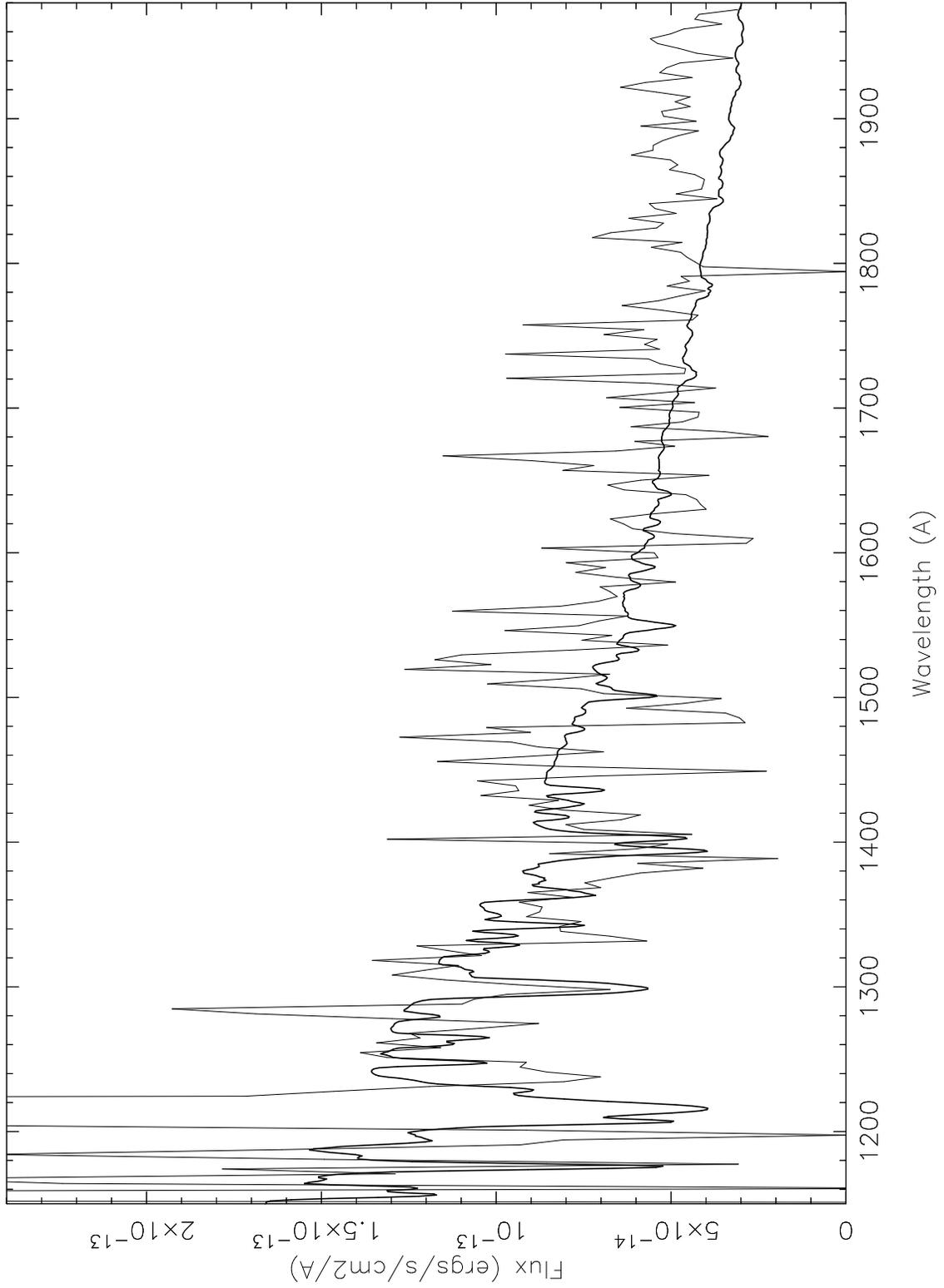}
\caption{Plot of flux versus wavelength for the spectrum SWP35239 
of the old nova Q Cygni. The solid curve is the best-fitting white dwarf model (see text for details).
}
\end{figure} 
\clearpage

\begin{figure} 
\epsscale{.90} 
\plotone{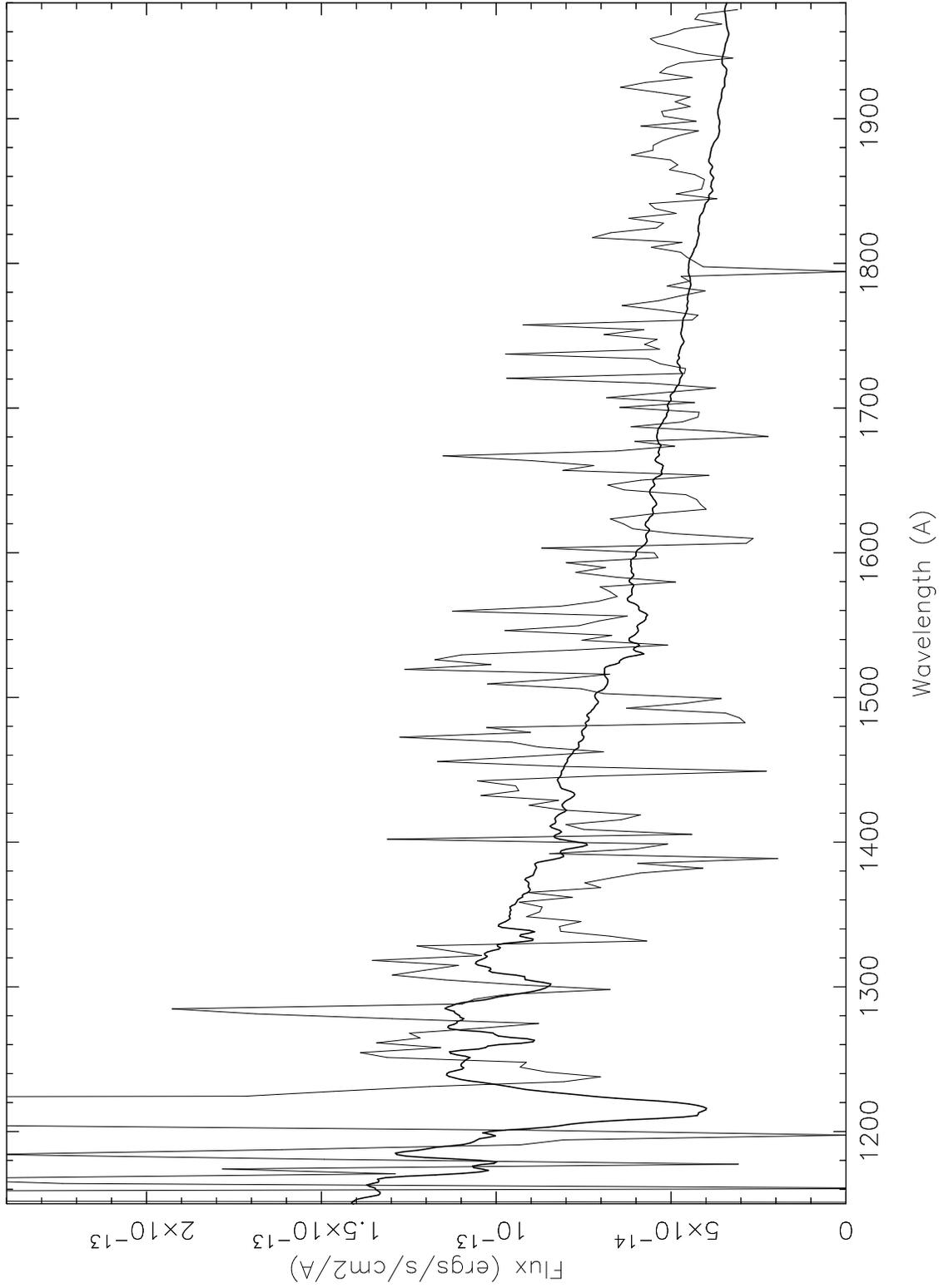}
\caption{The flux versus wavelength plot for the spectrum SWP35239 of the Q Cygni.
The solid curve is the best-fitting
accretion disk model (see text for details).
}
\end{figure}
 
\clearpage
\begin{figure} 
\epsscale{.90} 
\plotone{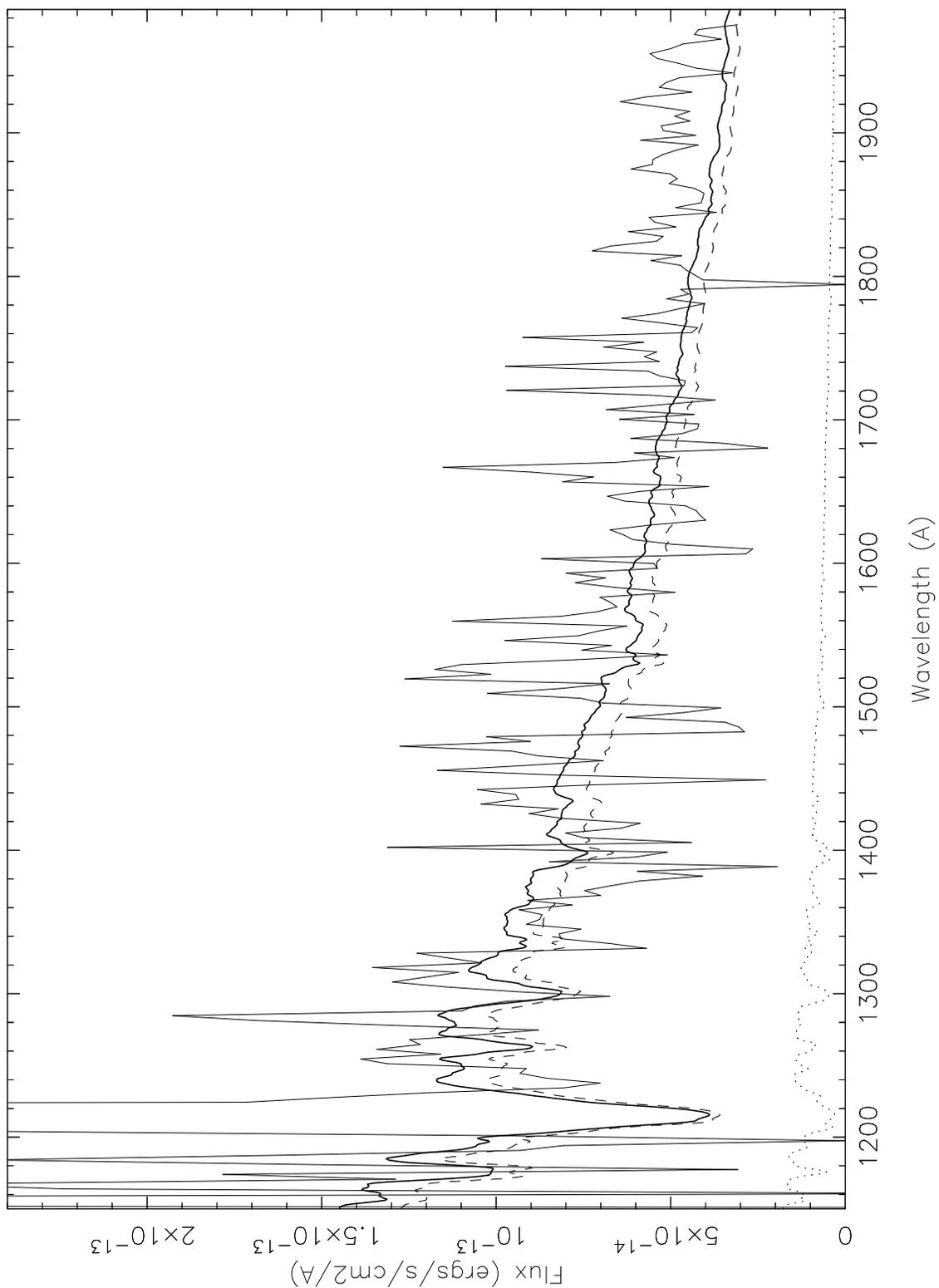}
\caption{The best-fitting combination accretion disk + white dwarf photosphere
synthetic fluxes to the spectrum SWP35239 of Q Cygni. The top solid curve is the
best-fitting combination, the dotted curve is the contribution of the white dwarf 
alone and the dashed curve is the accretion disk synthetic spectrum alone (see text for details).
}
\end{figure}

\end{document}